\documentclass[10pt,aip,cha,twocolumn,amsmath,showpacs,reprint]{revtex4-1}

\usepackage{graphicx}
\usepackage{amsfonts}

\graphicspath{{./}{../}{figures/}}

\usepackage{pstricks,pst-grad}

\newrgbcolor{lightviolet}{0.8 0.3 0.7}
\newrgbcolor{blue}{0 0 1}

\begin{document}

\title{Two-Particle Circular Billiards Versus Randomly Perturbed One-Particle Circular Billiards}

\author{Sandra Rankovi\'{c}}
\affiliation{Institute for Theoretical Physics, ETH Z\"urich, 8093 Z\"urich, Switzerland}
\affiliation{Mathematical Institute, University of Oxford, Oxford, OX1 3LB, UK}

\author{Mason A. Porter}
\affiliation{OCIAM, Mathematical Institute, University of Oxford, Oxford, OX1 3LB, UK}



\begin{abstract}

We study a two-particle circular billiard containing two finite-size circular particles that collide elastically with the billiard boundary and with each other.  Such a two-particle circular billiard provides a clean example of an ``intermittent" system. This billiard system behaves chaotically, but the time scale on which chaos manifests can become arbitrarily long as the sizes of the confined particles become smaller.  The finite-time dynamics of this system depends on the relative frequencies of (chaotic) particle-particle collisions versus (integrable) particle-boundary collisions, and investigating these dynamics is computationally intensive because of the long time scales involved.  To help improve understanding of such two-particle dynamics, we compare the results of diagnostics used to measure chaotic dynamics for a two-particle circular billiard with those computed for two types of one-particle circular billiards in which a confined particle undergoes random perturbations.  Importantly, such one-particle approximations are much less computationally demanding than the original two-particle system, and we expect them to yield reasonable estimates of the extent of chaotic behavior in the two-particle system when the sizes of confined particles are small. Our computations of recurrence-rate coefficients, finite-time Lyapunov exponents, and autocorrelation coefficients support this hypothesis and suggest that studying randomly perturbed one-particle billiards has the potential to yield insights into the aggregate properties of two-particle billiards, which are difficult to investigate directly without enormous computation times (especially when the sizes of the confined particles are small). 

\end{abstract}




\maketitle
   
%
%

\begin{quotation}

\textbf{A traditional \emph{billiard system} consists of a point particle confined in some domain (which is usually a subset of $\mathbb{R}^2$) and colliding perfectly elastically against the boundary of that domain.\cite{whatisbilliard,chernovbook}  Such billiards can have chaotic, regular (i.e., integrable), or mixed dynamics.\cite{AMSMushrooms,Mushrooms,Bunimovich}  For example, a finite-size circular particle confined within a circular boundary is integrable, but two circular particles confined in a circular domain yields chaotic dynamics, though regular behavior can persist for extremely long times.\cite{Steven}} 

\textbf{The chaotic dynamics in two-particle billiards appears via the dispersive mechanism \cite{whatisbilliard} as a result of particle-particle collisions, whereas particle-boundary collisions lead to regular dynamics when the boundary is circular. Consequently, although the long-time dynamics is chaotic, the regular transients can become arbitrarily long as one considers confined particles with progressively smaller radii.  It is desirable to find means to simplify investigations of the statistical properties arising from long-time transient dynamics in two-particle billiards, whose dynamics are not well understood and which require very long computations to simulate.  In this paper, we take a step in this direction by considering one-particle billiards with random perturbations and comparing diagnostics for measuring aggregate levels of chaotic dynamics in two-particle versus perturbed one-particle billiards. In particular, we consider two circular particles confined in a circular billiard and one-particle circular billiards with two types of random perturbations: one in which random perturbations are applied at times determined via a Poisson process and another in which random perturbations are applied at times given by actual particle-particle collision times from the two-particle system.}

\textbf{The two-particle circular billiard considered in the present paper provides a clean example of an ``intermittent" system. There continues to be considerable interest in intermittent billiards,\cite{bunimovichVela} and this example in particular deserves many future investigations.}  

\end{quotation}


\section{Introduction}
 
Systems of hard particles interacting via perfectly elastic collisions provide paradigm examples for studying the foundations of statistical mechanics.\cite{Dellago}  Among the primary examples used to study classical and quantum chaos in conservative systems are \emph{billiard systems},\cite{whatisbilliard,chernovbook} which can be implemented experimentally in both classical and quantum settings.\cite{chaosbook,MilnerBill,gutzwiller}  

Billiards are one of the most important types of Hamiltonian system.\cite{whatisbilliard} Typical classical Hamiltonian systems are neither fully chaotic nor fully regular (i.e., integrable) but instead have ``mixed" dynamics.\cite{chaosbook}  That is, their phase space has both regular and chaotic regions.  However, generic mixed systems are very difficult to analyze, so it is important to study Hamiltonian systems with simpler but non-generic mixed dynamics that allow more thorough analysis.\cite{Mushrooms,Steven,AMSMushrooms,Saito,debievre}  Billiard systems are among the most important systems for such pursuits.

The most commonly studied type of billiard system consists of a single point particle confined in a closed planar region. The particle collides against the boundary of the region such that its angle of incidence equals its angle of reflection.  More general types of billiards have also been investigated. For example, the study of open billiards, which contain a hole or leak through which particles can escape, has become increasingly prominent.\cite{AltmannLeak,DettmannOpen}  However, it is much less common---but nevertheless extremely interesting---to study few-particle billiards, in which a small number of finite-size confined particles collide elastically both against a billiard boundary and against each other.\cite{Steven}  The balls move freely between collisions.


Depending on the geometries of the billiard boundaries and the confined particles, it is possible for few-particle billiards to exhibit both regular and chaotic features.  This situation, however, is somewhat different from the idea of mixed dynamics mentioned above. Namely, although the dynamics of few-particle billiards are fully chaotic in the infinite-time limit, it is possible to construct systems such that integrable dynamics last for arbitrarily long periods of time.  In particular, this can occur as one considers confined particles with progressively smaller radii in two-particle billiard systems in which both the billiard balls and the billiard table are shaped like circles.\cite{Steven}  This example thereby provides a clean example of an ``intermittent" system.\cite{chaosbook}  Intermittent systems contain ``sticky" regions near which typical trajectories can get trapped for very long times and they can be notoriously difficult to study in detail.  Accordingly, it is useful to examine ``simple" (relatively speaking) examples of such systems, and studying intermittency in billiards provides a good avenue for such investigations.\cite{bunimovichVela}

A circular billiard table with a confined finite-size circular particle is integrable, and every particle-boundary collision in a two-particle circular billiard leads to regular dynamics.  However, particle-particle collisions in this two-particle system ultimately lead to chaotic dynamics (via the dispersive mechanism) in the infinite-time limit.  The dynamics of this system arise from a competition between the integrable particle-boundary collisions and the chaotic particle-particle collisions.  The relative frequency of the latter versus the former becomes smaller as the radii of the confined particles become smaller, and the integrable transients can therefore last for arbitrarily long times. Consequently, although the dynamics are eventually chaotic if one waits long enough, one must examine the transient dynamics to achieve a thorough understanding of this system.

Investigating the transient dynamics of two-particle billiards entails very long computation times, which is particularly true when the integrable transients are long. Accordingly, we use numerical computations to attempt to discern when a perturbed circular one-particle billiard can give a reasonable approximation for statistical properties of the two-particle billiard. When this is the case, which we show happens for small particle radii, one can attempt to gain insights into the original two-particle system by studying a perturbed one-particle system.

The rest of this paper is organized as follows. In Section \ref{sec2}, we describe the model of a circular billiard with confined circular particles of finite radius. For simplicity, we assume that both particles are of the same size. In Section \ref{sec3}, we describe some measures of the statistical properties of the system's dynamics. In Section \ref{sec4}, we compare these statistical properties for the two-particle circular billiard and for two types of perturbed one-particle circular billiards. 
In Section \ref{sec5}, we summarize our results.


\section{Modelling a Circular Billiard System} \label{sec2}

We study a two-particle billiard that occupies a region $Q \subset \mathbb{R}^2$ with a circular boundary $\partial Q$. Each of its confined, finite-size, circular particles of mass $m$ and normal momentum $p_{n}=mv_{n}$ moves freely until it encounters either the boundary or the other particle. To understand this system, we need to start from the simplest case. Accordingly, we assume in this paper that $m=\|v\|$ for both particles and also that they have the same radius. (We consider several different radii in our computations.) Without loss of generality, we set the radius of the billiard table to be $1$.  

The model that we consider is a simplification of a real-life billiard system, as we do not allow particles to change speeds when they collide with each other.
Allowing such a change in speeds would conserve momentum in situations in which the center-of-mass motion is initially nonzero. (When momentum is conserved, one can examine a two-particle billiard in two dimensions as a one-particle billiard in four dimensions.)  Our simplification simplifies the dynamics, but it retains many interesting features, and it is more tractable to study than the full system.

Consider the dynamics of one of the two particles in our system. When it collides against the boundary, its momentum changes according to
\begin{equation}\label{momchange}
 	\vec{p'} = \vec{p} - 2 \langle \vec{p},\vec{n} \rangle \vec{n}\,,
\end{equation}
where $\vec{p}$ is the momentum before the collision, $\vec{p'}$ is the momentum after the collision, and $\vec{n}$ is the unit normal to the boundary $\partial Q$ at the collision point. If the particles collide against each other instead, then the change in momentum of one particle still obeys equation (\ref{momchange}), but now $\vec{n}$ is the normal to the tangent of the boundary of the other particle at the point of collision.

In a dynamical system, a Poincar\'{e} section is a lower-dimensional subspace of phase space that is transversal to the flow. In a two-particle billiard, we can obtain a Poincar\'{e} section separately for each particle.  Consider the coordinates $(s_i,\sin(\theta_i))$, where $s_i \in [0,2\pi]$ is the arclength coordinate of the boundary at a collision between particle $i$ and the boundary and $\theta_i\in [0,2\pi]$ is the corresponding angle of collision against the boundary.\cite{chaosbook}  We construct a Poincar\'e section by tracking the coordinate values of such collisions for each particle while throwing away the continuous flow and particle-particle collisions in between such particle-boundary collisions.  This yields a four-dimensional Poincar\'e section, as two dimensions from phase space are eliminated by only considering the locations of particle-boundary collisions.  (We lose one dimension from using a Poincar\'e section, and we lose a second dimension via energy conservation.)
 
\begin{figure}[h!]
\centering
\includegraphics[width=.9\columnwidth]{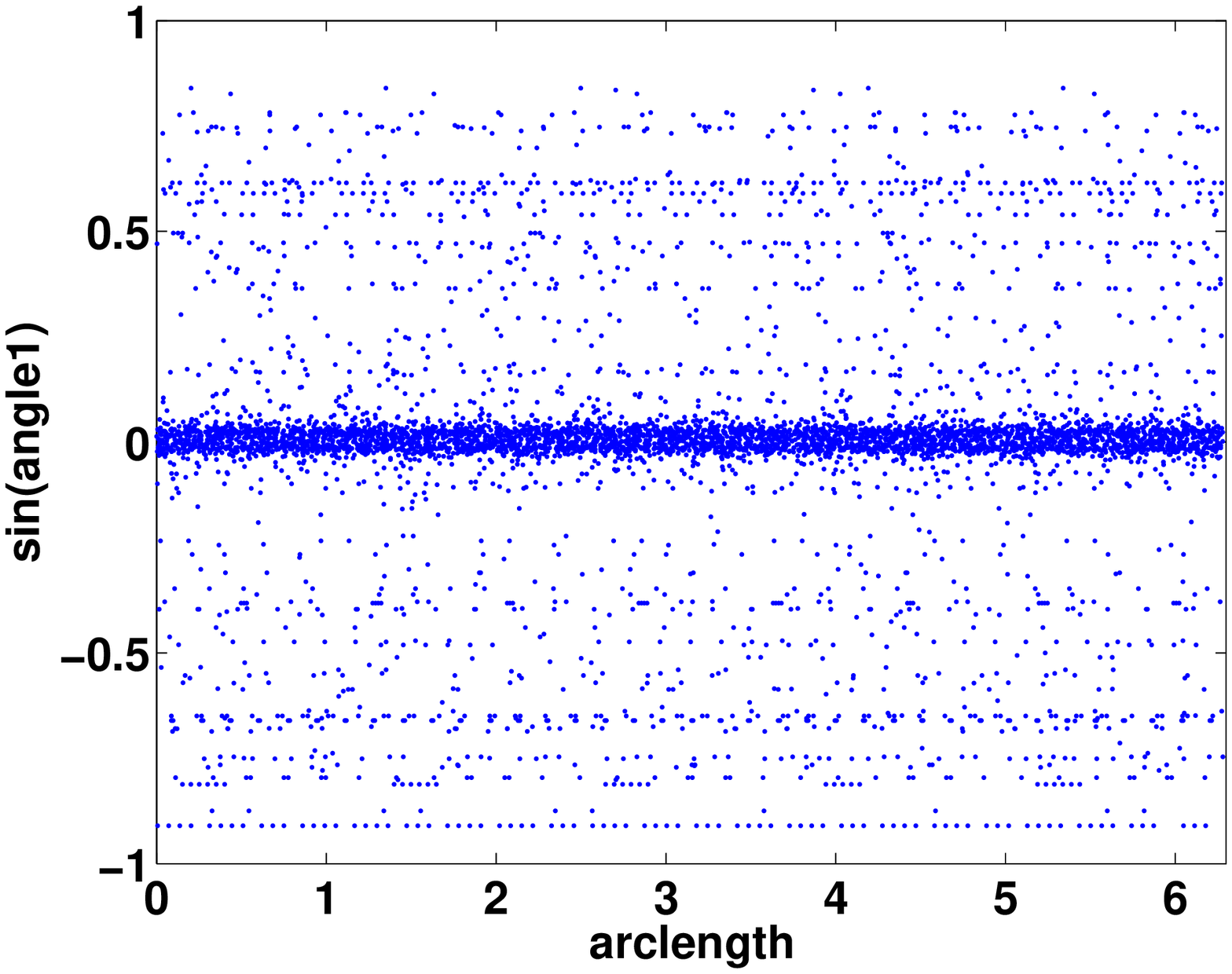} 
\includegraphics[width=.9\columnwidth]{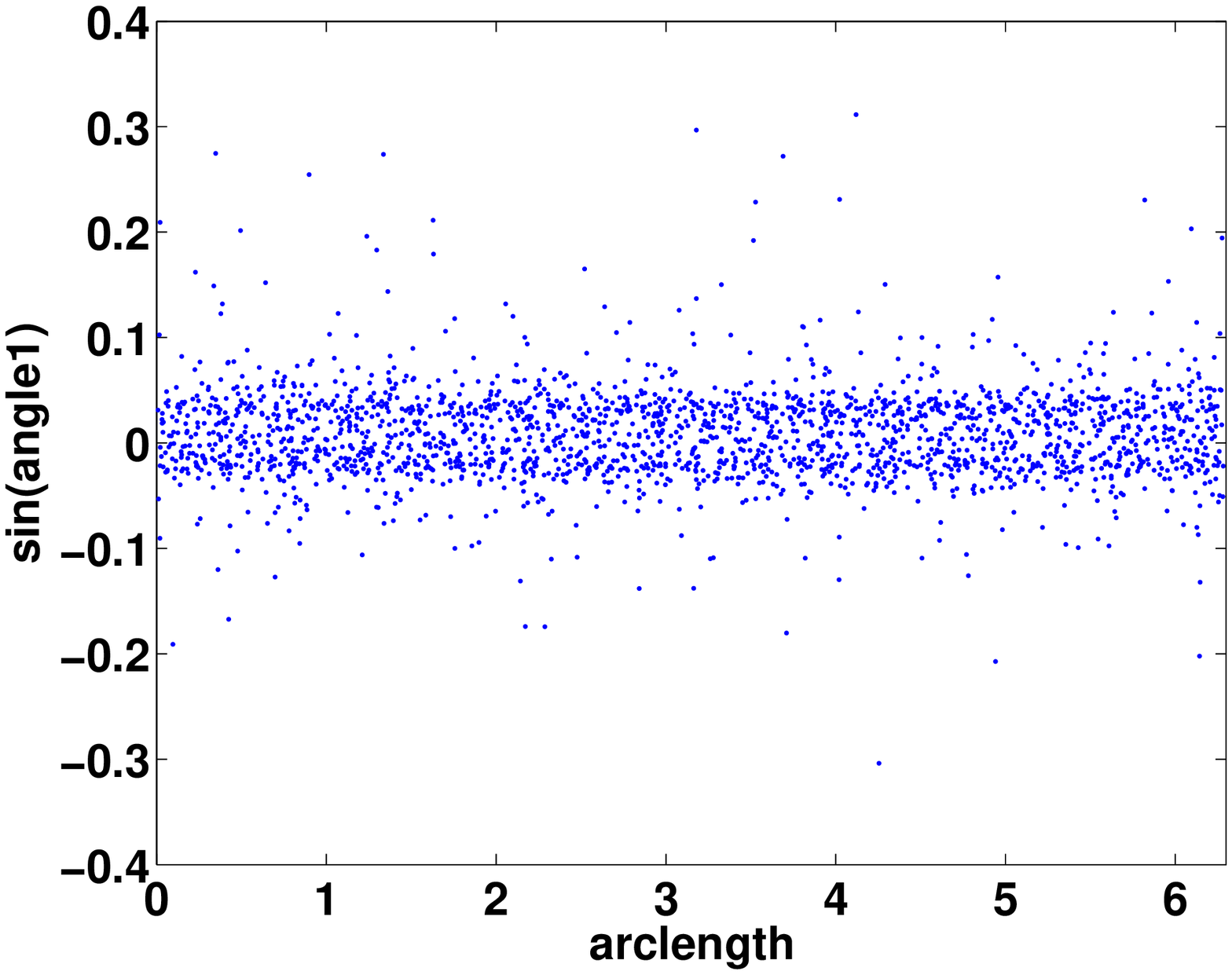} 
\caption{(Color online) Projections of example Poincar\'{e} sections for one particle in a two-particle circular billiard. We use the same initial conditions for both panels.  (As discussed in the main text, we choose initial conditions uniformly at random.)  In panel (a), the radius of each of the the confined particles is $0.02$, and we can observe fairly regular behavior (horizontal segments from consecutive particle-boundary collisions and parts of circles from recurrent behavior).  In panel (b), the radius of each confined particle is $0.014$. The behavior now appears to be more chaotic, as there are more isolated points and fewer regular features. 
}
\label{fig:Poincare3}
\end{figure}

In Fig.~\ref{fig:Poincare3}, we show two-dimensional projections of examples of Poincar\'{e} sections for one of the particles in the two-particle system. We consider examples in which each confined particle has a radius of $0.02$ (top panel) or $0.014$ (bottom panel). We plot the dependence of $\sin(\theta_i)$ of particle-boundary collisions versus the boundary location (i.e., the arclength $s_i$) of the collision.  To construct Fig.~\ref{fig:Poincare3}, we use discrete-time simulations with 15000 time steps in which each time step has a duration $\tau$ of up to 0.3 time units of the simulation (and to exactly 0.3 time units when there are no particle-particle or particle-boundary collisions during the time step).  We give a more precise description of our simulation algorithm below. When the radii of the confined particles are small, as in these two examples, the dynamics can exhibit regimes of regular-looking behavior even for long computation times.

Whether one can discern transient regular dynamics in the Poincar\'e sections depends on the initial conditions, computation time, and the sizes of the confined particles.  In particular, regions of regular behavior (which appear as incomplete lines and circles) are more evident in the top panel then in the bottom panel, although there are regions of regular behavior in the latter plot as well. Such regular features arise from particle-boundary collisions, which are angle-preserving. Horizontal segments represent consecutive particle-boundary collisions with the same angle, and parts of circles represent regions of recurrent behavior. Chaotic features arise from particle-particle collisions, whose dispersive character results in changes to $\theta_i$ in subsequent particle-boundary collisions and leads to divergence of trajectories (and the splattering of dots in Poincar\'e sections).

In these simulations, we determine initial conditions as follows.  For each of the two confined particles, we choose initial distances between its center and the center of the billiard table uniformly at random from $(0,1)$.  As the particles have finite radii (i.e., they are not point particles), we discard any configuration in which the particles overlap with each other or any particle overlaps with the billiard boundary. We do this by checking the overlapping condition and choosing a new random initial position until there is no overlap with the other particle or the boundary. We choose the initial angle (i.e., velocity) of each particle uniformly at random from the interval $[0,2\pi)$. Because the magnitude of the speed for each particle is always equal to $1$, this gives the starting velocity of each particle.  

We use a uniform distribution to determine our ensembles of initial conditions because it is the simplest choice.  It would be interesting to repeat our computational experiments for other distributions and to compare the results obtained for different choices.


In this paper, we perform computations for many values for the radius $r = r_1=r_2$ of the confined particles.  For a given family of computations, we use the same ensemble of initial conditions for each choice of $r$.  For a given family, we thus choose initial conditions (as described above) using the largest employed value of $r$.  This guarantees that there is never any particle-particle or particle-boundary overlap.  As with our choice of using a uniform distribution, we have made this choice because of its simplicity: when there is no overlap between the particles at the largest radius, then there is also no overlap for the smaller values of $r$ and we can therefore obtain precisely the same set of initial conditions for each value of $r$.


Our algorithm for simulating a two-particle billiard proceeds as follows. After each discrete time step (of fixed duration $\tau$), we calculate the positions and velocities of both particles.  For each time step, we check if a collision occurs (either between the two particles or between a particle and the billiard boundary).  If no collisions occur, we determine the new particle positions based on their free movement after a time $\tau$ (the velocities are unchanged).  Otherwise, the particles evolve freely for the (shorter) time step $\tau' < \tau$ until the next collision, and we then calculate their new positions and velocities immediately after that collision.



\section{Quantifying Chaotic Dynamics}\label{sec3}

Although the Poincar\'e sections in Fig.~\ref{fig:Poincare3} illustrate regular features that result from particle-boundary collisions and chaotic dynamics that result from particle-particle collisions, it is desirable to quantify the statistical properties resulting from the interplay of these two types of events as functions of the sizes of the confined particles. As diagnostics to describe chaotic dynamics in our system, we use recurrence-rate coefficients, autocorrelation coefficients, and finite-time Lyapunov exponents.


\subsection{Recurrence Plots and Recurrence Rates}

\textit{Recurrence plots (RPs)} were introduced as a tool to visualize recurrences in a variable $x_{i}$ in phase space.\cite{RP,Eckmann} Recurrence of states is a typical feature of 
chaotic systems, and it is traditional to try to find them by visualizing high-dimensional phase spaces as projections to two-dimensional or three-dimensional spaces.  RPs take a different approach and yield a visualization using an $N \times N$ matrix, where $N$ denotes the number of time steps in a simulation.

To construct an RP, we start with the formula\cite{RP}
\begin{equation}\label{rpform}
	R_{{i,j}}=\Theta(\varepsilon-\| x_{i}-x_{j}\|)\,,  
\end{equation}	
where $x_{i} \in \mathbb{R}^{m}$ and $i,j \in \{1,\ldots,N\}$.  Additionally, $\varepsilon$ is a threshold distance, $\| \cdot \|$ is a norm, and $\Theta( \cdot )$ is the Heaviside function.  In this paper, we always use the $L_{2}$-norm.\cite{Zbilut2002}

We compute RPs for the temporal evolution of the positions of the confined particles. We choose one of the two particles in the system and construct its recurrence plot. The variable $x_{i}$ represents the 
position coordinates of the chosen particle inside the billiard at time $i$. We are interested in how close and how often a particle returns to 
a given position $(x,y)$ that it has visited previously in its trajectory. An RP includes a dot in the location $(i,j)$ if the positions of the particle center at times $i$ and $j$ are within a threshold distance $\varepsilon$ from each other. The threshold should be small enough so that the particle can be considered to have approached sufficiently close to the previously visited locations, but it should also be large enough to keep computations reasonably efficient.\cite{MarwanEncounters}
 
 
In this paper, we present results using the threshold value $\varepsilon=0.001$.
(Computations using the value of $\varepsilon=0.01$ gave similar qualitative results.)
Early RP studies suggested that one should choose $\varepsilon$ to be a few percent of the phase-space diameter,\cite{Mindlin1992} and similar comments have been made about problems with circular symmetry.\cite{RP} Additionally, some authors have used an $x_i$-dependent threshold $\varepsilon_i$,\cite{Eckmann} but we have elected to use a uniform value. The constant value for the threshold yields a symmetry in the recurrence plot: 
the RP always has a main diagonal and is symmetric about it.

\begin{figure}[h!]
\centering
\includegraphics[width=.95\columnwidth]{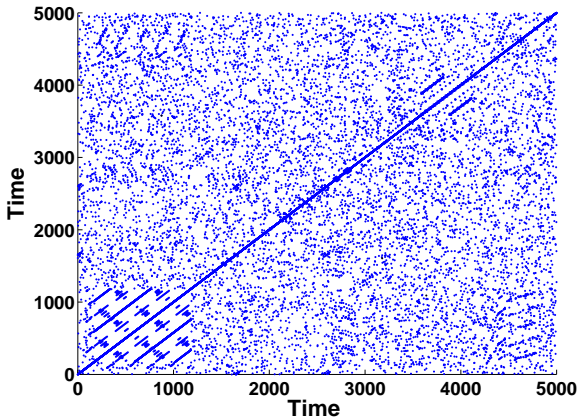} 
\caption{(Color online) Example recurrence plot with Euclidean norm for a two-particle circular billiard with two confined circular particles (each of which has a radius of $0.02$).
}
\label{fig:RecPlot}
\end{figure}

In Fig.~\ref{fig:RecPlot}, we show an example RP for one particle in a two-particle circular billiard. As with Poincar\'e sections, splatters of dots illustrate irregular behavior. The lines parallel to the diagonal represent regions of regular behavior, in which trajectories visit the same region of position space at different times. The isolated points by themselves don't contain any information about the system. However, the occurrence of isolated points next to the lines in an RP can be an indication of chaotic dynamics,\cite{RP} which we know occurs in this system. The lengths of the diagonal segments are determined from the durations of regular dynamics.\cite{RPAltmann}

An RP can be used to define a recurrence-rate ($RR$) coefficient \cite{Eckmann,RP}
\begin{equation}
	RR(\varepsilon)= \frac{\sum_{i,j=1}^N (R_{{i,j}})}{N^2}\,,
\end{equation}
which measures the density of recurrence points in an RP and can be used as a measure of complexity in a dynamical system. As the discrete time $N\rightarrow \infty$, the coefficient $RR(\varepsilon)$ represents the probability for a state to recur in an $\varepsilon$-neighborhood in 
position space.\cite{RP}

\subsection{Autocorrelation}

\textit{Autocorrelations} measure the tendency for observations of the same system made at different time points to be related to one another.\cite{gardiner} That is, an autocorrelation describes a correlation in a time series with respect to its own past and future values. A negative autocorrelation value indicates that the direction of influence is changing as a function of time, whereas a positive autocorrelation value can be construed as a tendency for a system to remain in a similar state from one observation to the next. 

For a series $x$ consisting of $N$ observations, the autocorrelation coefficient $R_{l}$ calculated between the time series and the same series lagged by $l$ time units is given by the formula
\begin{equation}
	 R_{l} = \frac{\sum_{i=1}^{N-l}\left(x_{i}-\langle x\rangle\right)\left(x_{i+l}-\langle x\rangle\right)}{\sum_{i=1}^{N}\left(x_{i}-\langle x\rangle\right)^2}\,,
\end{equation}
where $\langle x \rangle$ is the mean value of $x$ among the $N$ observations.

In this paper, we will compute autocorrelation coefficients for time series of angles of particle-boundary collisions.


\subsection{Lyapunov Exponents}

The largest \textit{Lyapunov exponent} $\lambda$ measures the rate of separation of trajectories in a dynamical system.\cite{Strogatz}
It is defined using the equation
 \begin{equation}
 	\|\delta d(t)\|=e^{\lambda t}\|\delta d(0)\|\,, 
\end{equation}	
where $\delta d(t)$ is the time-$t$ separation of two trajectories that start a distance $\delta d(0)$ apart. We use a finite-time version of the Lyapunov exponent because of the finite-time nature of numerical computations.  The dynamics of a system are only predictable up to the \emph{Lyapunov time}, which is defined as the time it takes for two neighboring trajectories to diverge by a distance equal to $e$.

A positive Lyapunov exponent indicates that trajectories separate from each other exponentially fast. For chaos to manifest, there needs to be a positive Lyapunov exponent and trajectories also need to mix.
A positive Lyapunov exponent implies that there is a local instability, and mixing implies that  trajectories of individual particles get arbitrarily close to each other arbitrarily often if the system evolves for a sufficiently long time.\cite{EckmannRuelle,Strogatz,chaosbook}

Our two-particle system behaves chaotically in the
$t \rightarrow \infty$ limit,\cite{Steven} though its most interesting behavior occurs during the potentially extremely long transients, and we use numerical computations to study such dynamics.  To calculate the largest finite-time Lyapunov exponent, we use Benettin's algorithm,\cite{Benettin1976} which assumes that small perturbations in initial conditions will stretch primarily along the most unstable direction in phase space after a sufficiently long time. As discussed previously, we evolve the billiard system for $n$ time steps of duration up to $\tau$. (Recall that the duration is exactly $\tau$ if there is no collision, and it is the time $\tau' < \tau$ that elapses until the next collision if there is a collision.) The time step $\tau$ thus gives the unit of time for our numerical simulations. This yields a finite-time Lyapunov exponent of
\begin{equation}
 	\lambda_{\rm{\max}}(n)=
 \dfrac{1}{\tau n}\sum_{k=1}^{n}
 \left(\log\left(\dfrac{\|\delta d(k\tau)\|}{\|\delta d(0)\|}\right)\right)\,,
 \label{lambda}
\end{equation}
where $\|\delta d(k \tau)\|/\|\delta d(0)\|$ is the stretching factor due to an initial perturbation of size $\|\delta d(0)\|$. After each of the $n$ time steps (which we index by $k$), we evaluate $\log(\|\delta d(k \tau)\|/\|\delta d(0)\|)$ and compute $\lambda_{\rm{max}}$ as a mean over these $n$ evaluations.

Each of the simulations that we report in this paper\cite{Note2} uses either $n = 25000$ steps of duration $\tau = 0.2$ or $n = 15000$ steps of duration $\tau = 0.3$.
The size of the initial perturbation is always $\|\delta d(0)\| = 1/5000$. In our numerical computations, we report values for the largest finite-time Lyapunov exponent $\lambda_{\rm{\max}}(n)$ that is a mean over results that we obtain using an ensemble of initial conditions ($200$ or $1000$ in the examples shown). We choose these initial conditions using the procedure that we discussed in Section \ref{sec2}.

We calculate finite-time Lyapunov exponents by separately evaluating the stretching factor for the 
horizontal position variables
and vertical position variables
and then taking the maximum of the two corresponding exponents [$\lambda^x_{\rm{\max}}(n)$ and $\lambda^y_{\rm{\max}}(n)$] calculated using equation (\ref{lambda}). Note that we evaluate the stretching factors using position variables only rather than using all phase-space coordinates, so we are actually calculating a variant of Lyapunov exponents.

 

 


\section{Comparison of Two-Particle Billiards and Perturbed One-Particle Billiards} \label{sec4}

Because analytical calculations are very difficult for two-particle billiards, we use numerical computations to determine when statistical properties for the two-particle circular billiard can be approximated effectively by those for a perturbed one-particle billiard.  Perturbed one-particle billiards have the potential to be more tractable analytically than two-particle billiards, and they can require significantly less computational time (and fewer computational resources).

In this paper, we compare our two-particle billiard system with perturbed one-particle billiards using numerical simulations to compute the diagnostics (largest finite-time Lyapunov exponents, recurrence-rate coefficients, and autocorrelation coefficients) that we discussed in Section \ref{sec3}.

We consider two different types of perturbation to one-particle circular billiards, where we apply perturbations (``kicks") to the confined circular particle.  We first examine a perturbed one-particle billiard with time intervals between kicks determined using the exponential probability distribution. This corresponds to using a Poisson process to determine the number of perturbations that a particle experiences in a given amount of time. As we will explain in Section \ref{poissonsec}, we estimate the parameter for the exponential distribution (and hence for the Poisson process) from the statistics of the original two-particle system. The second type of perturbed one-particle circular billiard that we consider is one with random kicks at times that are determined directly from particle-particle collisions in the original two-particle billiard.  


\subsection{Random Perturbations in a One-Particle Billiard} \label{velocity} 
 
We apply random perturbations to a one-particle circular billiard at times determined by a Poisson process (see the discussion below) or directly from particle-particle collision times in the two-particle billiard. We impose the random perturbations by determining a velocity angle uniformly at random from the interval $[0,2\pi)$. We have chosen this type of perturbation because we are considering small confined particles, so particle-particle collisions are not very common, and we are assuming that we have no knowledge of the angle at which such a collision occurs. We have also considered angles chosen from the standard Gaussian distribution $\mathcal{N}(0,1)$. This approach appears to yield similar qualitative results, so we only show the results of our computations for the uniform distribution.
 

\subsection{Poisson Process for Determining Perturbation Times} \label{poissonsec}

Our original system consists of two circular particles of the same size confined inside a radius-1 circular billiard. These particles collide elastically against the boundary of the billiard and against each other.  As we discussed in Section \ref{sec2}, we choose the initial positions and velocities (i.e., angles) of the particles uniformly at random for the two-particle billiard. 
For our initial conditions in perturbed one-particle systems, we take the set of initial conditions for one of those two particles.  We compute each of our diagnostics for both two-particle billiards and one-particle billiards for each initial condition, and we show plots with the means of those diagnostics over all initial conditions.  (We consider as many as 1000 initial conditions for some calculations.)

When applying kicks in the perturbed one-particle billiards, we change the particle velocity (i.e., its angle) as explained in Section \ref{velocity}. 
Because of the randomness of the initial data, we assume that all of the time intervals between particle-particle collisions (and hence between particle perturbations in the one-particle systems) have the same distribution. Each such collision is thus independent from other collisions.  Note that we make this assumption for simplicity, and it is desirable to relax it. 

An approach that is similar to ours was used successfully in work by Dahlqvist et al.~\cite{dahlqvist1996} and by Baladi et al.,\cite{baladiEckmann} who investigated recurrence properties of intermittent dynamical systems using a probabilistic independence assumption about recurrence times.  Baladi et al. found that asymptotic properties can be influenced significantly by the tail of distributions (which, in the present context, is the distribution of particle-particle collision times).  Heavy tails are a hallmark of intermittency, so it is good to keep this observation in mind for the problem that we study.

We seek a means to estimate kick times for a particle in a perturbed one-particle billiard to attempt to obtain an approximation for some statistical properties of the original two-particle billiard without having to simulate the original two-particle system. 

\begin{figure}[h!]
\centering
\includegraphics[width=.95\columnwidth,bb = -160 151 773 640]{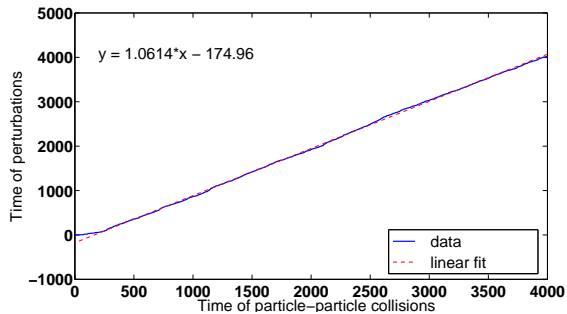} 
\caption{(Color online) Comparison of the times of particle-particle collisions in a two-particle circular billiard with Poisson-process kick times in a perturbed one-particle circular billiard.  We determined the initial conditions in our simulations uniformly at random, and we consider confined particles of radius $r = r_1 = r_2 = 0.008$.
}
\label{fig:Times}
\end{figure}

Because of our independence assumption, we use a Poisson process to describe the sequence of particle-particle collisions in our two-particle system. (One can, of course, use more sophisticated distributions based on a system's dynamics, and this is an important idea for future investigations.) The Poisson process was developed to model events that occur by chance and independently from each other while maintaining a constant intensity (i.e., the expected number of events per unit time is constant).\cite{Blank,Feller} The Poisson process in our example has a rate of ${1}/{\mu}$, where $\mu$ gives the mean (continuous) amount of time per particle-particle collision.  Because the times of kicks of the confined particle in a perturbed one-particle billiard should correspond to the times for the particle-particle collisions in the original two-particle system, a Poisson process gives estimates of these perturbation times. In Fig.~\ref{fig:Times}, we show a plot of the perturbation times determined by a Poisson process in a one-particle system versus the original times of the particle-particle collisions in the two-particle billiard. 
In this example, the radius of each confined particle is $r = r_1 = r_2 = 0.008$. We fit a linear function to this curve using the method of least squares. The high quality of our fit 
suggests that using a Poisson process to determine perturbation times is a reasonable approximation.

It follows from the definition of a Poisson process that the time intervals between the perturbations are given by independent and identically distributed (IID) random variables. Because the system behaves ergodically in the infinite-time limit, such a distribution for particle-particle collisions seems plausible.  The sequence of particle-particle collisions in the two-particle billiard and hence of the perturbations in the one-particle billiard then behaves as a Markov process, so the future of the process depends only on the present state (and is independent of the past). However, when considering finite-time dynamics, recurrences in the system lead to deviations from the Poisson-process approximation to the collision times, and this becomes increasingly noticeable as the radii of the confined particles become larger.  We note, moreover, that computations of escape rates in circular billiards with a hole\cite{DettmannOpen} suggest that the above Poisson picture is too simplistic.  Examining this in more detail is a very interesting idea for future work.

\begin{figure}[h!]
\centering
\includegraphics[width=.95\columnwidth]{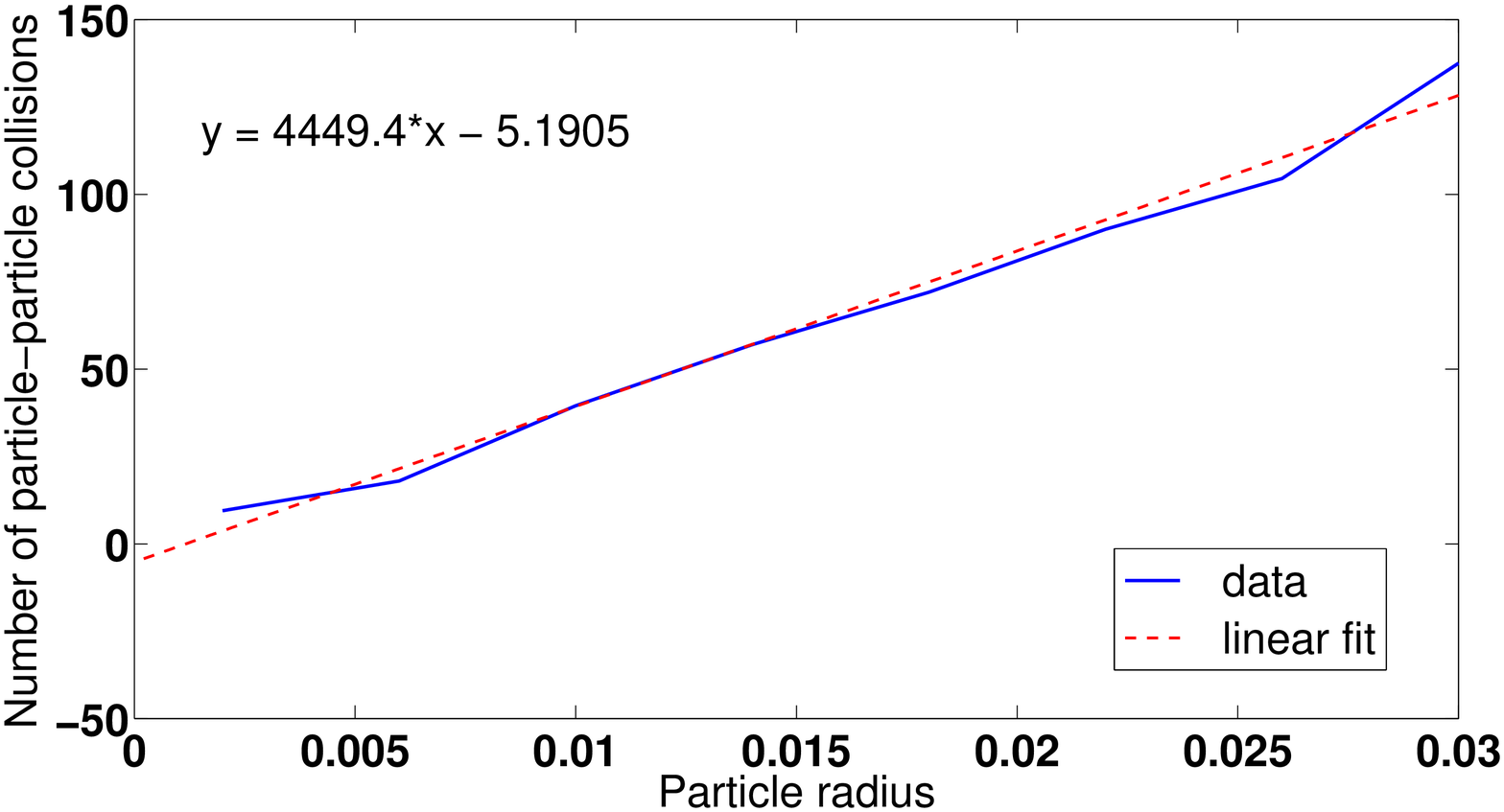}
\caption{(Color online) Mean number of particle-particle collisions in a circular two-particle billiard as a function of the radius $r = r_1 = r_2$ of the confined particles.
} 
\label{fig:ballHit}
\end{figure}

To avoid having to simulate the original two-particle billiard to obtain the mean number of particle-particle collisions (and hence the rate of the Poisson process) that we will subsequently use in a perturbed one-particle billiard, we do a set of simulations of the two-particle billiard first to obtain estimates of this quantity. In Fig.~\ref{fig:ballHit}, we show the dependence of the number of particle-particle collisions on the particle radius in the two-particle billiard for the smaller particle sizes that we considered. We fit a linear curve to this graph using the method of least squares, and the fit is reasonably good. This suggest that, for a given particle radius, one can pre-determine the rate of its associated Poisson process and then use that parameter value when investigating a perturbed one-particle billiard.  However, we did not do this when comparing our results for two-particle billiards and perturbed one-particle billiards.  Instead, to obtain better estimates of the Poisson rates, we perform numerical simulations of the original two-particle system for each initial condition.

\subsection{Results of the Comparisons} 
  
In this section, we compute diagnostics for circular two-particle billiards (which is computationally intensive) and compare them to the same diagnostics computed for perturbed one-particle billiards. For it to be reasonable to apply random perturbations to the velocity variable in one-particle billiards, we need to consider long computations and small confined particles.  For simplicity, we assume that both confined (circular) particles in the two-particle billiard have the same size, though of course it would be interesting to examine more general situations. We consider several different particle radii in order to examine the effect of particle size on the diagnostics.  As discussed in Section \ref{sec3}, the diagnostics that we use to measure chaotic dynamics are the maximal finite-time Lyapunov exponent, the recurrence-rate coefficient, and the autocorrelation coefficient for the angle of particle-boundary collisions.

\begin{figure}[h!]
\centering 
\includegraphics[width=.95\columnwidth]{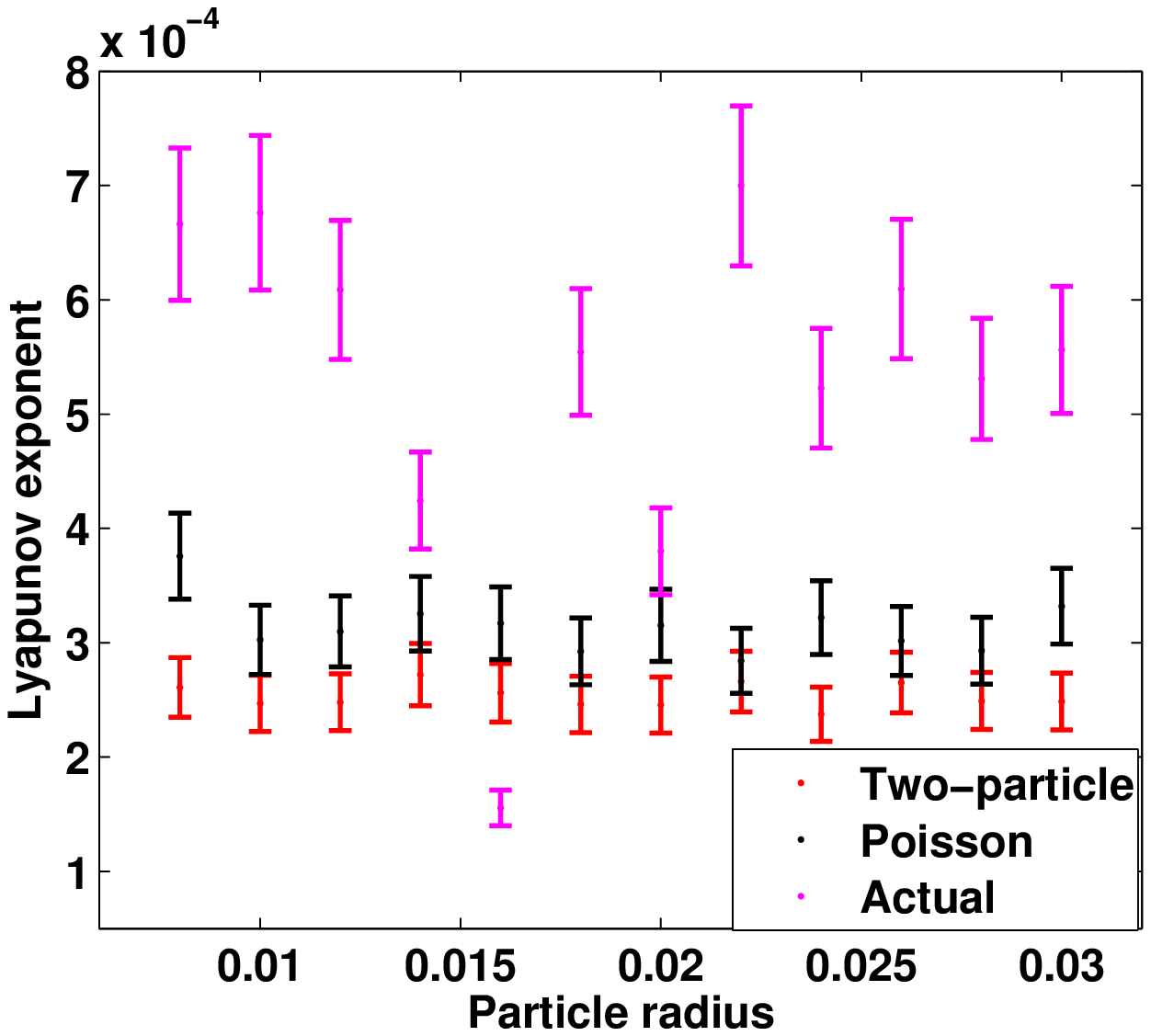}
\includegraphics[width=.95\columnwidth]{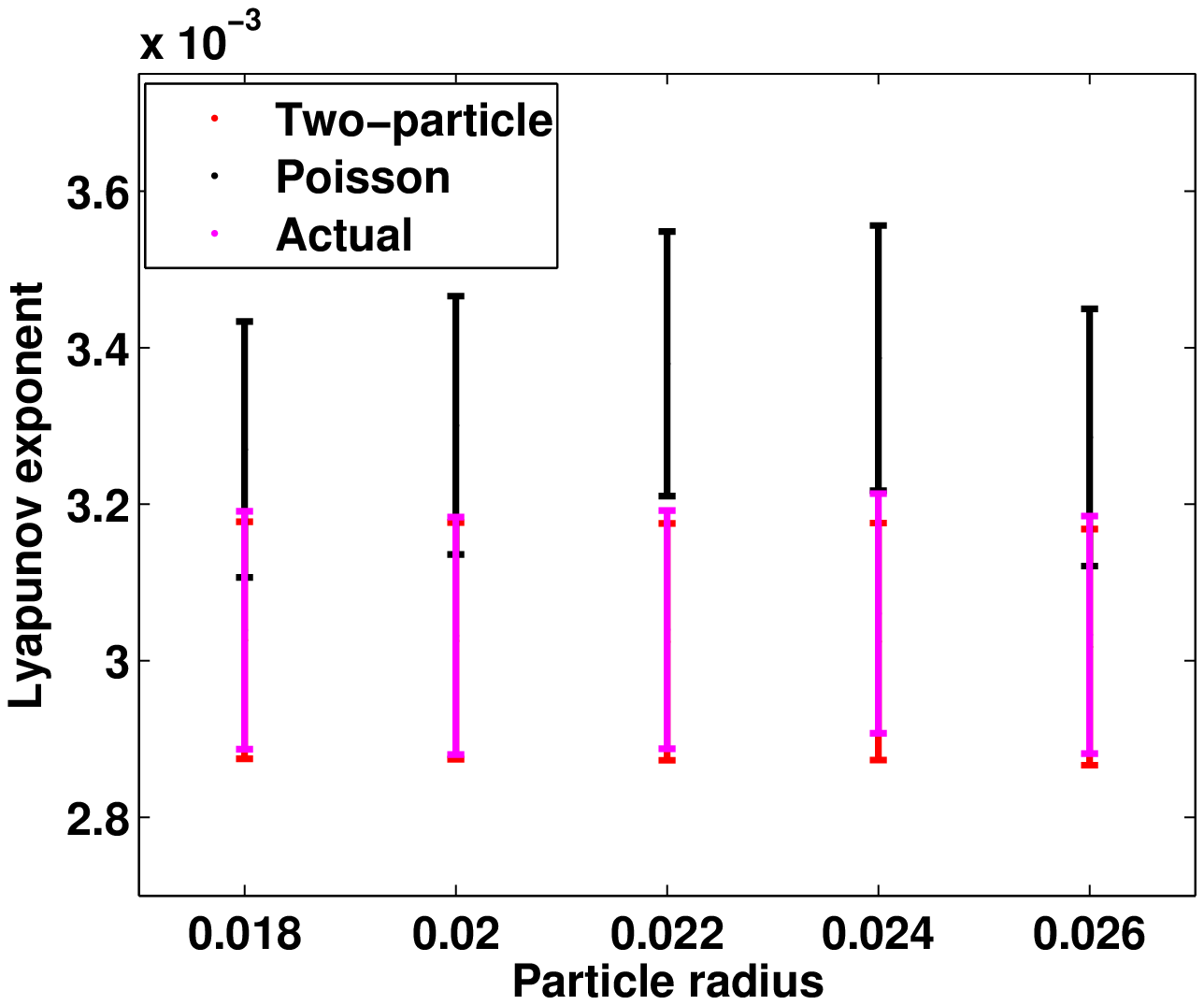}
\caption{(Color online) Finite-time Lyapunov exponents for a two-particle circular billiard (`Two Particle'), a perturbed one-particular circular billiard with perturbation times drawn from a Poisson distribution (`Poisson'), and a perturbed one-particle circular billiard with perturbation times drawn from the times of particle-particle collisions in the two-particle system (`Actual') as a function of particle radius for (identical) confined particles.  In the top plot, we consider radii ranging from $0.008$ to $0.03$.  For each radius, we average over 200 random initial conditions (see the discussion in the text for how we choose these initial conditions) and simulate for 25000 time steps (of 0.2 time units each). In the bottom plot, we consider radii ranging from $0.008$ to $0.026$. For each particle radius, we average over 1000 random initial conditions and simulate for 15000 time steps (of 0.3 time units each). In the top plot, observe that our computations for the Poisson-time perturbed one-particle billiard gives a good approximation to those for the two-particle billiard.  The Lyapunov exponents from the actual-time perturbed one-particle billiard have the same order of magnitude as the other two sets of values. In the bottom plot, observe that our calculations for the actual-time perturbed one-particle billiard gives an excellent approximation to those for the two-particle billiard.  The Lyapunov exponents from the Poisson-time perturbed one-particle system are also very similar. The mean values are located in the centers of the error bars, which depict 1/10 of a standard deviation above and below the mean.
}
\label{fig:Lyap1000}
\end{figure}

\begin{figure}[h!]
\centering
\includegraphics[width=.95\columnwidth]{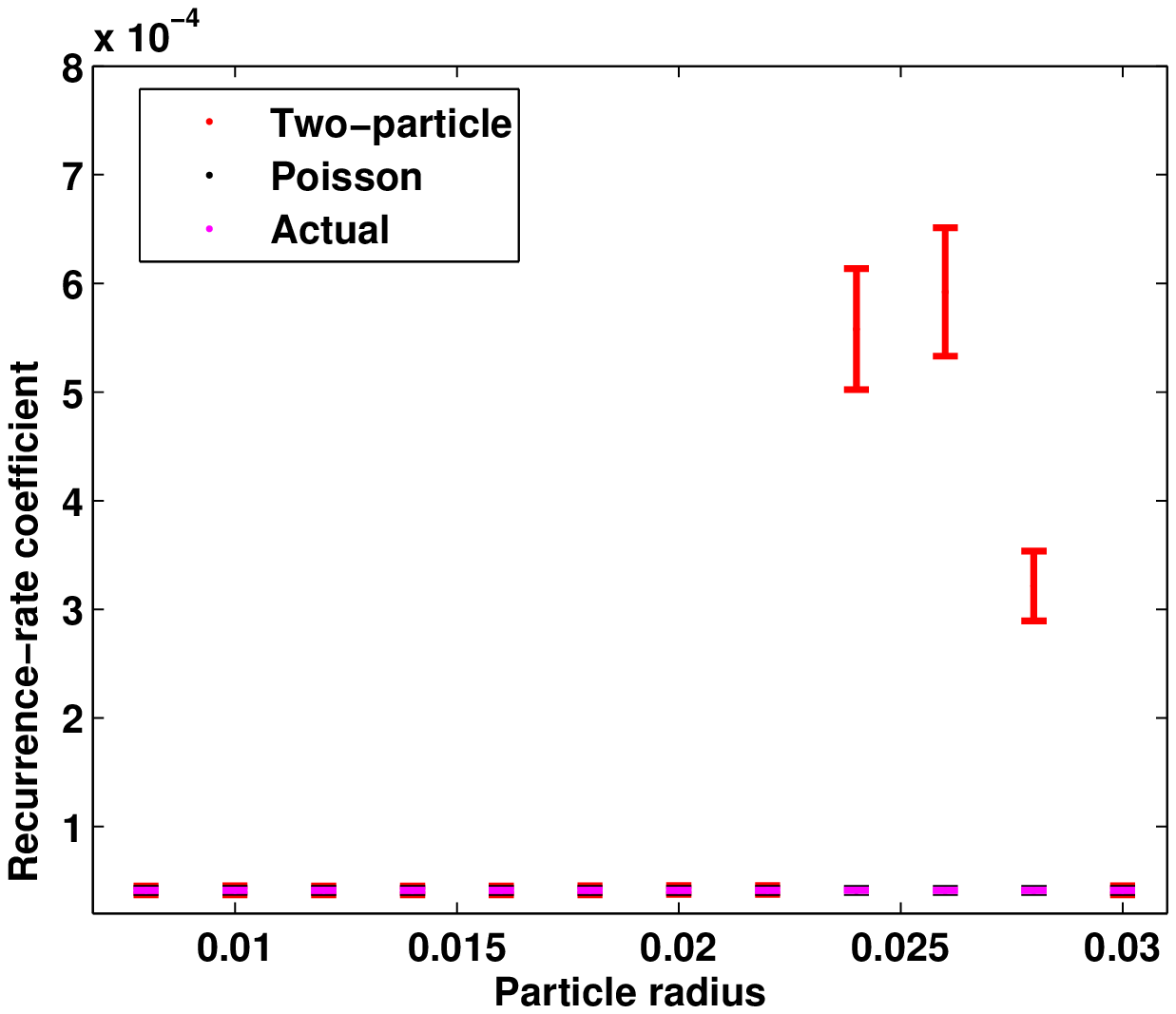} 
\includegraphics[width=.95\columnwidth]{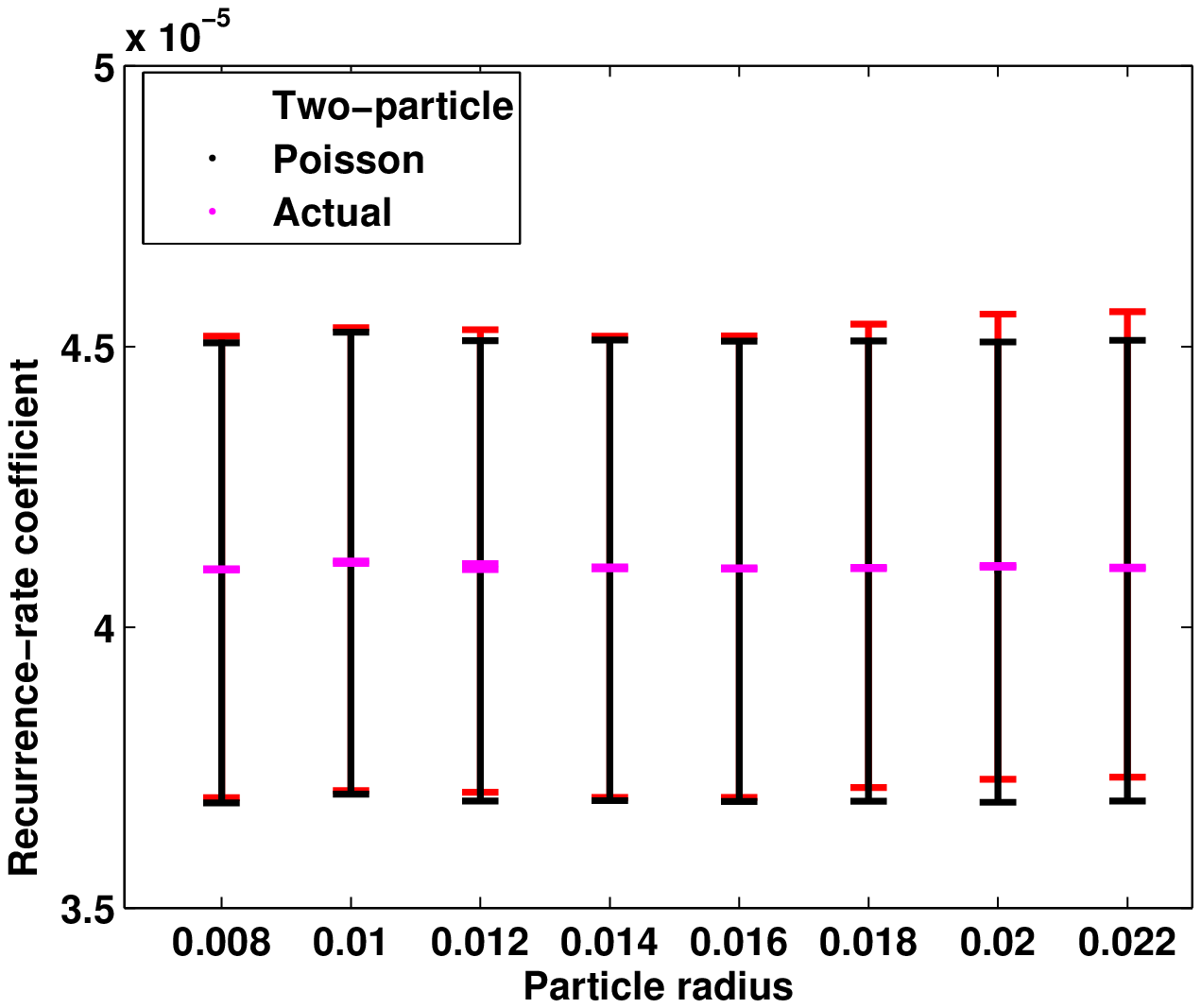} 
\caption{(Color online) Recurrence-rate coefficients for a two-particle circular billiard (`Two Particle'), a perturbed one-particular circular billiard with perturbation times drawn from a Poisson distribution (`Poisson'), and a perturbed one-particle circular billiard with perturbation times drawn from the times of particle-particle collisions in the two-particle system (`Actual') as a function of particle radius for (identical) confined particles with radii ranging from $0.008$ to $0.03$.  For each particle radius, we average over 200 random initial conditions and simulate for 25000 time steps (of 0.2 time units each). In the bottom plot, we show results only for small radii to make it easier to see the very good agreement between all three computations in that regime.  For larger radii, the two perturbed one-particle billiards still have similar RR coefficients, but the RR coefficient for the two-particle billiard exhibits an interesting balloon (depicted in the top plot) before shrinking again.  Even during the balloon, all three computations give values with the same order of magnitude. The mean values are located in the centers of the error bars, which depict 1/10 of a standard deviation above and below the mean.
} 
\label{fig:RR2}
\end{figure}

\begin{figure}[h!]
\centering
\includegraphics[width=.95\columnwidth]{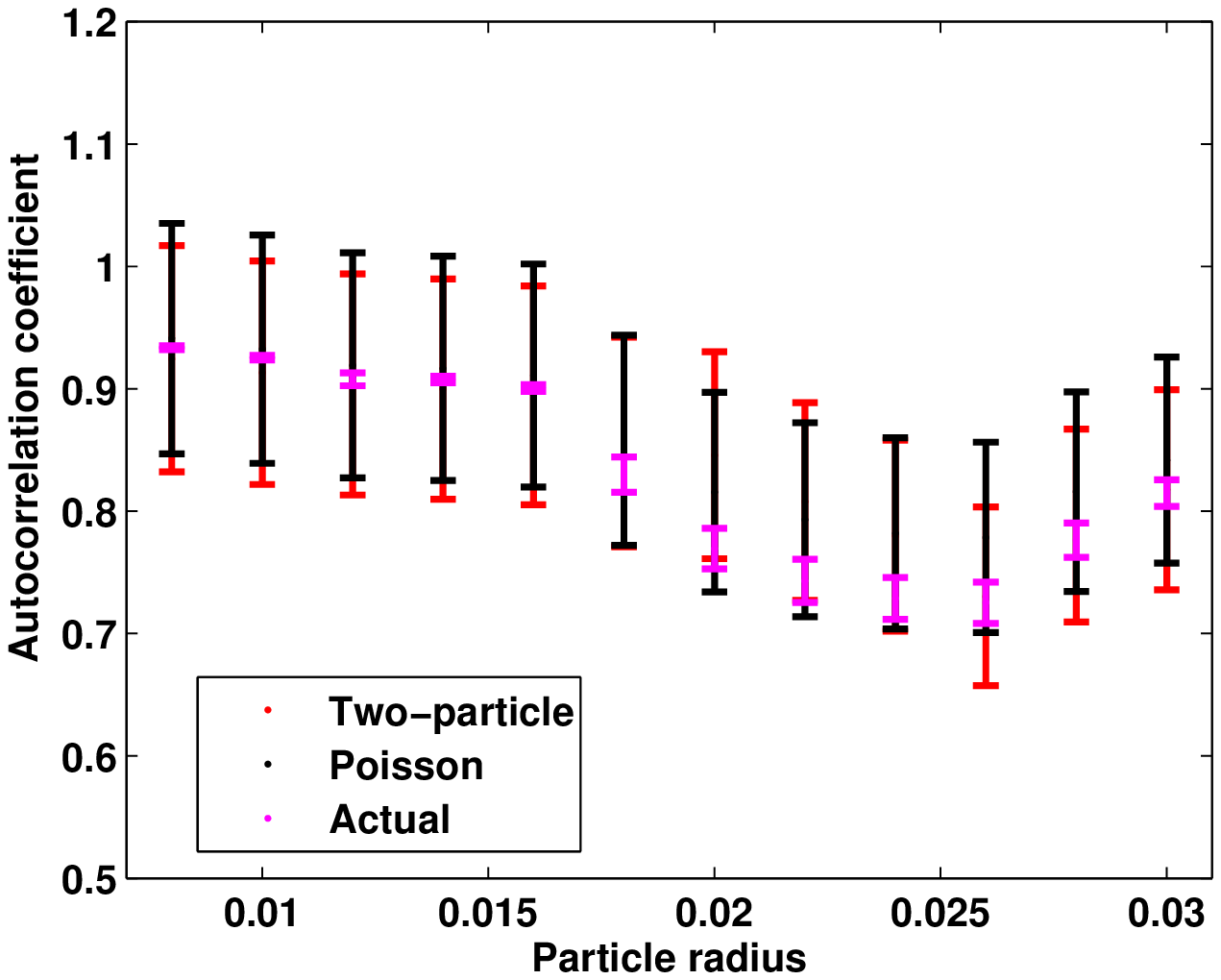} 
\includegraphics[width=.95\columnwidth]{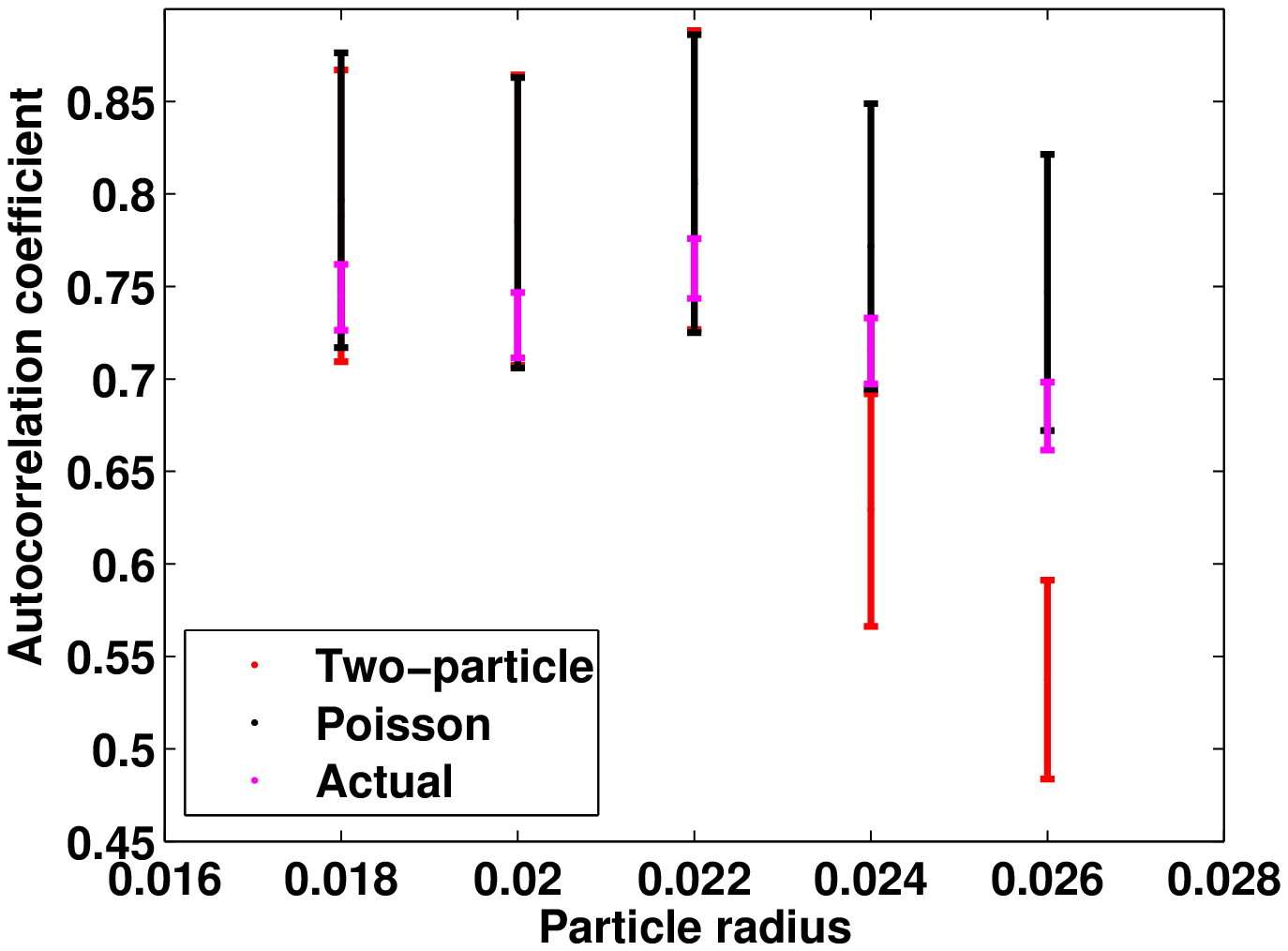} 
\caption{(Color online) Autocorrrelation coefficients for the angle of particle-boundary collisions for one particle in a two-particle circular billiard (`Two Particle'), a perturbed one-particular circular billiard with perturbation times drawn from a Poisson distribution (`Poisson'), and a perturbed one-particle circular billiard with perturbation times drawn from the times of particle-particle collisions in the two-particle system (`Actual') as a function of particle radius for (identical) confined particles. In the top plot, we consider radii ranging  from $0.008$ to $0.03$.  For each radius, we average over 200 initial conditions and simulate for 25000 time steps (of 0.2 time units each). In the bottom plot, we consider radii ranging from $0.008$ to $0.026$. For each radius, we average over 1000 initial conditions and simulate for 15000 time steps (of 0.3 time units each).  We observe excellent agreement for all three systems in the top plot. We still find good agreement in the bottom plot, though for larger radii the results for the two-particle billiard start to deviate quantitatively from those for the perturbed one-particle billiards. The mean values are located in the centers of the error bars, which depict 1/10 of a standard deviation above and below the mean.
} 
\label{fig:Corr200}
\end{figure}

In Figs.~\ref{fig:Lyap1000}--\ref{fig:Corr200}, we show the dependence on the radius of the confined particles of the diagnostics applied to one particle in our three systems (which we label as `Two-particle', `Poisson', and `Actual'). As discussed previously, we average each diagnostic over a large number of initial conditions.
In the plots, we also include error bars whose length is given by 1/5 of the standard deviation corresponding to the calculated mean value (i.e., the error bars show 1/10 above and below the mean).  The computed standard deviations are large because we choose our initial conditions uniformly at random and the systems are chaotic.  

As the plots illustrate, the diagnostics have values of the same order of magnitude for both the two-particle billiard and the perturbed one-particle billiards.  In some cases, these values are sometimes also very close to each other quantitatively. In general, the values of the diagnostics appear to become more dissimilar as one considers confined particles of larger radii, though we do not observe a monotonic dependence.
  
From our numerical computations alone, it is difficult to discern when our one-particle systems provide  ``good" approximations to the original two-particle billiard. However, we believe that examining two-particle billiards versus associated perturbed one-particle billiards has the potential to be valuable.  Additionally, these perturbed one-particle billiards are interesting systems to investigate in their own right.

Interestingly, drawing the perturbation times from a Poisson process sometimes yields better results than taking perturbation times from the actual times of particle-particle collisions in the original two-particle billiard, even though doing this uses less information from the original system. This result, together with Fig.~\ref{fig:Times}, suggests that it is sometimes reasonable to use a Poisson process to estimate the number of particle-particle collisions in the two-particle circular billiard.

\section{Conclusions and Discussion} \label{sec5}

Circular two-particle billiards provide a clean example of intermittent dynamics, so studying them in detail should be very useful for obtaining a better understanding of intermittency.  In the present paper, we have taken a step in this direction by comparing the properties of a circular two-particle billiard with two different circular one-particle billiard systems undergoing random perturbations.  We considered random perturbations at times determined in two different ways: (1) times taken from a Poisson distribution, and (2) times obtained directly from the times of particle-particle collisions in the original two-particle system.  

For such approximations to be reasonable, one should consider small confined particles, which entails very long computation time for the two-particle system, as particle-particle collisions occur much less frequently than particle-boundary collisions.  Consequently, the two-particle system exhibits long transients with regular behavior, although the system behaves chaotically in the $t\rightarrow\infty$ limit.  Accordingly, it can be very helpful to find situations in which it is reasonable to compare some of the aggregate properties of two-particle billiards to properties in perturbed one-particle systems (which require much less exhaustive simulations to study). We focused in the present paper on numerical simulations, but ideally perturbed one-particle billiards will also be studied analytically.

We considered Lyapunov exponents, recurrence-rate coefficients, and autocorrelation coefficients and found that the two-particle billiard and perturbed one-particle billiards have values of the same order of magnitude when the confined particles are small.  At times, we also found quantitative agreement. Because these diagnostics are used widely to measure the amount of chaos in a system, this suggests that the aggregate levels of chaos in the one-particle systems are similar to that in the original two-particle system.

Importantly, the one-particle approximations are much less computationally demanding than the original two-particle system, and we expect that they will also provide a much easier setting for analytical investigations, which are very difficult for two-particle billiards.  We thus propose that randomly perturbed one-particle billiards have the potential to provide insights on the aggregate (and intermittent) dynamics of associated two-particle billiards.


\section*{Acknowledgements}

We thank Leonid Bunimovich, Carl Dettmann, Steven Lansel, and two anonymous referees for helpful comments.  We are particularly grateful to Leonid Bunimovich for suggesting that we examine this problem from this point of view and to Carl Dettmann for several useful discussions and helpful comments on manuscript drafts. SR also thanks the Nuffield foundation and St John's College for financial support during the time of the research, Charles Batty for recommendation letters supporting the applications for the aforementioned funding, and the administration of Dartington House for office space provided during the research.  SR was affiliated with University of Oxford for most of the duration of this project.



\end{document}